\documentclass[longauth, traditabstract]{aa} 
\usepackage{graphicx}
\usepackage{txfonts}
\usepackage{natbib}
\begin{document}

\title{\textit{Herschel} observations of deuterated water towards Sgr~B2(M)\thanks{\textit{Herschel} is an ESA space
    observatory with science instruments provided by European-led
    Principal Investigator consortia and with important participation
    from NASA.}}

\author{
C.~Comito,\inst{1}
P.~Schilke,\inst{1,2}
R.~Rolffs, \inst{1,2}
D.~C.~Lis,\inst{3}
A.~Belloche,\inst{1}
E.~A.~Bergin,\inst{4}
T.~G.~Phillips,\inst{3}
T.~A.~Bell,\inst{3}
N.~R.~Crockett,\inst{4}
S.~Wang,\inst{4}
G.A. Blake\inst{3}
E.~Caux,\inst{5,6}
C.~Ceccarelli,\inst{7}
J.~Cernicharo,\inst{8}
F.~Daniel,\inst{8,9}
M.-L.~Dubernet,\inst{10,11}
M.~Emprechtinger,\inst{3}
P.~Encrenaz,\inst{9}
M.~Gerin,\inst{9}
T.~F.~Giesen,\inst{2}
J.~R.~Goicoechea,\inst{8}
P.~F.~Goldsmith,\inst{12}
H.~Gupta,\inst{12}
E.~Herbst,\inst{13}
C.~Joblin,\inst{5,6}
D.~Johnstone,\inst{14}
W.~D.  Langer,\inst{12}
W.D.~Latter,\inst{15} 
S.~D.~Lord,\inst{15}
S.~Maret,\inst{7}
P.~G.~Martin,\inst{16}
G.~J.~Melnick,\inst{17}
K.~M.~Menten,\inst{1}
P.~Morris,\inst{15}
H.~S.~P. M\"uller,\inst{2}
J.~A.~Murphy,\inst{18}
D.~A.~Neufeld,\inst{19}
V.~Ossenkopf,\inst{2,20}
J.~C.~Pearson,\inst{12}
M.~P\'erault,\inst{9}
R.~Plume,\inst{21}
S.-L.~Qin,\inst{2}
S.~Schlemmer,\inst{2}
J.~Stutzki,\inst{2}
N.~Trappe,\inst{18}
F.~F.~S.~van der Tak,\inst{20}
C.~Vastel,\inst{5,6}
H.~W.~Yorke,\inst{12}
S.~Yu,\inst{12}
M.~Olberg,\inst{22}
R.~Szczerba,\inst{23}
B.~Larsson,\inst{24}
R.~Liseau,\inst{22}
R.~H.~Lin,\inst{12}
L.~A.~Samoska,\inst{12}
E.~Schlecht\inst{12}
}
\institute{Max-Planck-Institut f\"ur Radioastronomie, Auf dem H\"ugel 69, 53121 Bonn, Germany\\
\email{ccomito@mpifr.de}
\and I. Physikalisches Institut, Universit\"at zu K\"oln,
              Z\"ulpicher Str. 77, 50937 K\"oln, Germany
\and California Institute of Technology, Cahill Center for Astronomy and Astrophysics 301-17, Pasadena, CA 91125 USA
\and Department of Astronomy, University of Michigan, 500 Church Street, Ann Arbor, MI 48109, USA 
\and Centre d'\'etude Spatiale des Rayonnements, Universit\'e de Toulouse [UPS], 31062 Toulouse Cedex 9, France
\and CNRS/INSU, UMR 5187, 9 avenue du Colonel Roche, 31028 Toulouse Cedex 4, France
\and Laboratoire d'Astrophysique de l'Observatoire de Grenoble, 
BP 53, 38041 Grenoble, Cedex 9, France.
\and Centro de Astrobiolog\'ia (CSIC/INTA), Laboratiorio de Astrof\'isica Molecular, Ctra. de Torrej\'on a Ajalvir, km 4
28850, Torrej\'on de Ardoz, Madrid, Spain
\and LERMA, CNRS UMR8112, Observatoire de Paris and \'Ecole Normale Sup\'erieure, 24 Rue Lhomond, 75231 Paris Cedex 05, France
\and LPMAA, UMR7092, Universit\'e Pierre et Marie Curie,  Paris, France
\and  LUTH, UMR8102, Observatoire de Paris, Meudon, France
\and Jet Propulsion Laboratory,  Caltech, Pasadena, CA 91109, USA
\and Departments of Physics, Astronomy and Chemistry, Ohio State University, Columbus, OH 43210, USA
\and National Research Council Canada, Herzberg Institute of Astrophysics, 5071 West Saanich Road, Victoria, BC V9E 2E7, Canada 
\and Infrared Processing and Analysis Center, California Institute of Technology, MS 100-22, Pasadena, CA 91125
\and Canadian Institute for Theoretical Astrophysics, University of Toronto, 60 St George St, Toronto, ON M5S 3H8, Canada
\and Harvard-Smithsonian Center for Astrophysics, 60 Garden Street, Cambridge MA 02138, USA
\and  National University of Ireland Maynooth. Ireland
\and  Department of Physics and Astronomy, Johns Hopkins University, 3400 North Charles Street, Baltimore, MD 21218, USA
\and SRON Netherlands Institute for Space Research, PO Box 800, 9700 AV, Groningen, The Netherlands
\and Department of Physics and Astronomy, University of Calgary, 2500
University Drive NW, Calgary, AB T2N 1N4, Canada
\and Chalmers University of Technology, SE-412 96 G\"oteborg, Sweden, Sweden
\and N. Copernicus Astronomical Center, Rabianska 8, 87-100, Torun, Poland
\and Department of Astronomy, Stockholm University, SE-106 91 Stockholm, Sweden}

\abstract{Observations of HDO are an important complement for studies of water, because they give strong constraints on the formation processes -- grain surfaces versus energetic process in the gas phase, e.g. in shocks.  The HIFI observations of multiple transitions of HDO in Sgr~B2(M) presented here allow the determination of the HDO abundance throughout the envelope, which has not been possible before with ground-based observations only. The abundance structure has been modeled with the spherical Monte Carlo radiative transfer code RATRAN, which also takes radiative pumping by continuum emission from dust into account.  The modeling reveals that the abundance of HDO rises steeply with temperature from a low abundance ({ $2.5\times 10^{-11}$}) in the outer envelope at temperatures below 100~K through a medium abundance ({ $1.5\times 10^{-9}$}) in the inner envelope/outer core, at temperatures between 100 and 200~K, and finally a high abundance ({ $3.5\times 10^{-9}$}) at temperatures above 200~K in the hot core. 
}

   \keywords{ISM: abundances --- ISM: molecules
               } 
   \titlerunning{Deuterated water in Sgr B2}
	\authorrunning{Comito et al.}
   \maketitle
%

\section{Introduction}\label{intro}

Water is known to be a fundamental ingredient of the interstellar medium.  It is a major coolant of star-forming clouds, but, apart from some high-lying maser transitions, is unobservable from the ground because of the atmospheric absorption.  The satellites SWAS and Odin \citep{Neufeld2003, Sandqvist2006} have observed the ground-state ortho-water line, and ISO has observed highly excited transitions \citep{Goicoechea2004, Cernicharo2006, Polehampton2007}, although with low spectral resolution. Water is also an important reservoir of oxygen and therefore a key ingredient in the chemistry of oxygen-bearing molecules. Investigations of the water abundance using many lines covering a wide range of excitation levels with high spectral resolution have been among the main reasons for building the HIFI instrument on board \textit{Herschel} \citep{deGraauw2010, Pilbratt2010}.

Water can be formed in the gas phase through an ion-molecule channel that is initiated by a reaction of H$^+$ or H$_3^+$ with atomic oxygen. Alternatively, water can also be formed through highly endothermic reactions that involve atomic oxygen, which reacts with molecular hydrogen to form OH, which in turn can react with molecular hydrogen to form H$_2$O.  The first process can work in rather diffuse clouds, where the ionization fraction is high, while the second one is efficient only in shocks, because only there the temperatures are high enough to overcome the high endothermicity. In dense gas, which is neither highly ionized nor very hot, these channels do not work efficiently.  There it is assumed that water is formed on grain surfaces, where oxygen can react with atomic hydrogen.  These icy grain mantles can be studied through observations of the water stretching bands in ice in the infrared \citep[e.g.,][]{Gibb2000}, by which the existence of water ice is a well established observational fact.  It is then assumed that this ice mantle is returned to the gas phase either through thermal heating (where water ice mantles are destroyed at about 100~K), or through shock sputtering.  The net result is that gas-phase water, irrespective of its formation channel, is often observed in regions where the energetics are such that it is not {\it a priori} clear if the water has a gas phase or grain surface origin.

Deuterium fractionation, i.e.\ the enrichment of the deuterated counterpart of a molecular species much above the cosmic [D]/[H] ratio, is a process that is caused by the zero-point energy difference of the vibrational potential. The deuterated species have a lower zero-point energy because of their larger reduced mass.  For H$_2$O and HDO, the difference in energy is 3.4 kJ/mole or 409~K. That means that deuterium fractionation for H$_2$O takes place mainly at low temperatures.  In other words, the [HDO]/[H$_2$O] ratio should be \emph{high} if the water in question has been formed at a low temperature, i.e.\ on cold grains, but \emph{low} if it is a product of either PDR or shock chemistry.  Observing the [HDO]/[H$_2$O] ratio is therefore an important clue to the formation mechanisms of H$_2$O.

This was the goal of the study by \citet[hereafter C03]{Comito2003}, who investigated the ground-state HDO line at 893~GHz observed with the Caltech Submillimeter Observatory towards Sgr~B2(M) and (N), as well as some published high-excitation HDO lines observed with the IRAM 30-m telescope \citep{Jacq1990}.  The HDO observations were compared with those of the o-H$_2^{18}$O ground-state line from SWAS \citep{Neufeld2003}, and of higher-energy p-H$_2^{18}$O transitions observed with the IRAM 30-m by \citet{Gensheimer1996}.  

The HDO and H$_2^{18}$O ground-state lines, observed in absorption, are formed in the outer envelope, while the high-energy transitions trace the central hot core. Up to now,  no observations tracing  the inner Sgr~B2(M) envelope/outer core were available. \citet{Comito2003} concluded that in the outer layer, H$_2$O was predominantly shock-produced, while the [HDO]/[H$_2$O] ratio in the hot core showed, as expected, a grain surface origin. Recent HIFI observations in the framework of the Guaranteeed Time Key Program \emph{Herschel Observations of Extra-Ordinary Sources (HEXOS)}  make it now possible to extend the study of the HDO distribution to the full envelope of Sgr~B2(M).

Sgr~B2(M) and its neighbor Sgr B2(N) are some of the most massive star-formation sites in our Galaxy. Understanding their history and structure  will help understanding the formation of massive Galactic center  clusters, such as Arches, and will also have ramifications for even more extreme star-formation processes in starburst galaxies, and even in the early universe.

 \begin{figure}[htbp]
   \centering
   \includegraphics[bb=40 15 535 400,width=0.95\columnwidth]{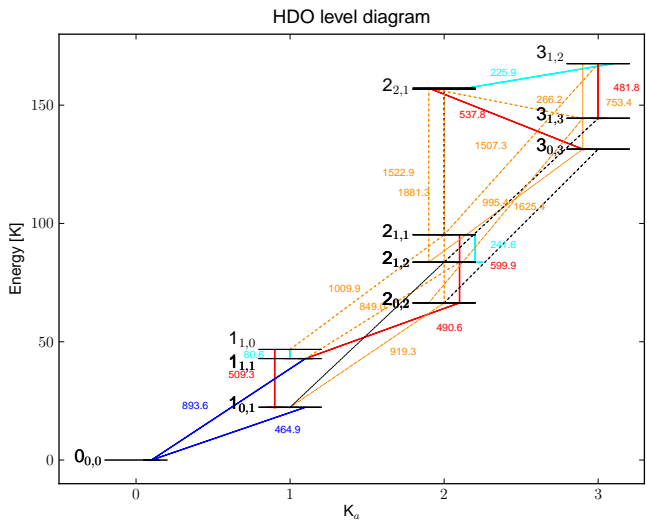}
  \caption{{ Level diagram of the HDO lines in the present dataset. The frequency of each transition is indicated in GHz. Transitions are color-coded as follows: in \emph{solid red}, HIFI observations in band 1a and 1b (this work); in \emph{dashed orange}, transitions falling within the overall HIFI range (to be observed within the {\it HEXOS} Key Program); in \emph{dark blue}, submillimeter-wavelength observations (C03; M. Gerin, priv. comm.); in \emph{light blue}, millimeter-wavelength observations from \citet{Jacq1990}, \citet{Nummelin1998},  and Belloche et al., in preparation \citep[cf.][]{Belloche2008}; in \emph{black}, transitions that fall outside of the HIFI range. }}\label{fig:HDO_levels}
\end{figure}

\section{Observations}

Full spectral
scans of HIFI bands 1a and 1b towards Sgr~B2(M) ($\alpha_{J2000} = 17^h47^m20.35^s$ and $\delta_{J2000} =
-28^{\circ}23'03.0''$) have been carried out on March 20101
 and 2, providing coverage of the frequency range 479 through
 637 GHz.  
 \begin{table}[htdp]
\caption{Observed HDO lines used in the modeling.  }\label{tab:lines}
\begin{center}
\begin{tabular}{rccrlc}
\hline
\hline
  & J$_K$  &  $\nu$ & E$_{\rm l}/k$  &  Telescope   &  Ref.  \\
   &              & (GHz)   &      (K)               &                        &   \\
\hline   
1 & $3_{1,2} - 3_{1,3}$                &  481.8  &  144.5             &   HIFI             & $a)$\\
2 & $2_{0,2}  - 1_{1,1}$                &  490.6   &  42.9             &    HIFI            & $a)$ \\
3 & $1_{1,0}  - 1_{0,1}$              &   509.2  &  22.3               &     HIFI           &  $a)$\\
4 & $2_{2,0} - 3_{0,3}$              &    537.8  &  131.4              &   HIFI              & $a)$\\
5 & $2_{1,1} - 2_{0,2}$              &    599.9 &    66.4              &    HIFI              & $a)$\\
\hline
6 &  $1_{1,0} - 1_{1,1}$             &  80.6       &  42.9               & IRAM 30-m   & $b)$\\
7 & $3_{1,2}  - 2_{2,1}$              &   225.9 &  156.7              &  IRAM 30-m      & $c)$\\
8 & $2_{1,1}  - 2_{1,2}$               &   241.6 &    83.6            & SEST                & $d)$ \\
9 & $1_{0,1}  - 0_ {0,0}$              &   464.9 &     0.0              & CSO                 & $e)$ \\
10 & $1_{1,1}  - 0_{0,0}$               &   893.6 &     0.0           & CSO                 & $f)$ \\
\hline
\end{tabular}
\end{center}
Rows 1 through 5: summary of all the HDO transitions observed in the HIFI bands 1a and 1b towards Sgr B2(M) (this work). Rows 6 through 9: list of previously observed HDO features that were used to constrain the results of our model (see Sect.~\ref{results}). References: $a)$, this work; $b)$, Belloche et al., in preparation; $c)$, \citet{Jacq1990}; $d)$, \citet{Nummelin1998}; $e)$, M. Gerin, priv. comm.; $f)$, \citet{Comito2003}.
\end{table}%
The HIFI Spectral Scans are carried out in Dual Beam Switch (DBS)
mode, where the DBS reference beams lie approximately 3$^{\prime}$
east-west or 6$^{\prime}$ apart. The wide band spectrometer (WBS) is used as a back-end,
providing a spectral resolution of 1.1 MHz over a 4-GHz-wide
Intermediate Frequency (IF) band.  A HIFI Spectral Scan consists
of a number of double-sideband (DSB) spectra, tuned at different Local
Oscillator (LO) frequencies, where the spacing between LO settings
is determined by the ``redundancy'' selected by the
observer \citep{Comito2002}. The molecular spectrum of Sgr~B2(M) in
band 1a and 1b has been scanned 
with a redundancy of 4 and 8, respectively, which means that every
frequency has been observed 4 and 8 times  respectively in each
sideband. Multiple observations of the same frequency at different LO
tunings are necessary to separate the lower-sideband (LSB) from the
upper-sideband (USB) spectra.

The data were calibrated through the standard pipeline distributed with
version 2.9 of HIPE \citep{Ott2010}, and subsequently exported to
CLASS\footnote{{\it Continuum and Line Analysis Single-dish Software},
  distributed with the GILDAS software, see http://www.iram.fr/IRAMFR/GILDAS.} using the HiClass task within HIPE. Deconvolution of
the DSB data into single-sideband (SSB) format was performed with
CLASS. All the HIFI data presented here, spectral features \emph{and}
continuum emission, are deconvolved SSB spectra. 
{ Both
horizontal (H) and vertical (V) polarizations were obtained. A discrepancy in flux of up to 20\% is observed between the two. 
Because the H data in band 1a and 1b show fewer instrumental artifacts than the V data,  we will here show
 H-polarization spectra only}. The intensity scale
is main-beam temperature, and is a results of applying a beam efficiency
correction of 0.69 for band 1a, and 0.68 for band 1b \citep{Roelfsema2010} to the deconvolved data  .

{ Table~\ref{tab:lines} and the level diagram in  Fig.~\ref{fig:HDO_levels} summarize the HDO transitions that were observed so far towards Sgr~B2(M). Together with those observed with HIFI (top five lines in Table~\ref{tab:lines}, red transitions in Fig.~\ref{fig:HDO_levels}), our dataset includes mm-wavelength transitions from the IRAM-30m telescope and from SEST, and submm-wavelength transitions acquired with the CSO (references are indicated in Tab.~\ref{tab:lines}).} The frequencies were obtained from the JPL Molecular Spectroscopy Catalog \citep{Pickett1998}. 

   %
   
%
 %

\section{Results}\label{results}
The spherical Monte Carlo radiative transfer code RATRAN
\citep{Hogerheijde2000} is employed to compute the intensity of the
molecular lines and of the dust continuum. The source model \citep[for a more detailed description see][]{Rolffs2010} corresponds to a centrally heated source with a density distribution that  follows a radial power law
with an index of 1.5. The best match between the observed and modeled continuum
is achieved  through the basic MRN grain size distribution \citep[][and references therein]{Ossenkopf1994}, that is, no formation of
ice mantles and no coagulation onto the grains is included. The assumed source luminosity is $6.3\times10^6$~L$_{\sun}$. At the inner radius, 1365 AU, the density is $1.5\times10^8$~cm$^{-3}$, and the temperature is 1400~K. The
temperature is computed in an approximate way from central
heating, using the diffusion equation in the inner part and the balance
between heating and cooling in the outer part, with an effective
temperature of the dust photosphere of 54.5~K. The temperature
has a steeper gradient in the inner part (200~K at 0.07~pc; 100~K at 0.14~pc) and decreases to
15~K at the radius of 12.3~pc, where the density is $2.7\times10^3$~cm$^{-3}$. From this point on, the temperature and density remain constant, until the outer envelope radius of 22.5~pc is reached \citep{Lis1989}.  This model yields a total H$_2$ column density of $7\times10^{24}$~cm$^{-2}$. { The HDO collisional rates have been provided by LAMBDA \citep[Leiden Atomic and Collisional Database, ][]{Schoeier2005}, based on the work of \citet{Green1989}.}

\begin{figure}
   \centering
   \includegraphics[bb=70 250 575 500,angle=-90,width=0.75\columnwidth]{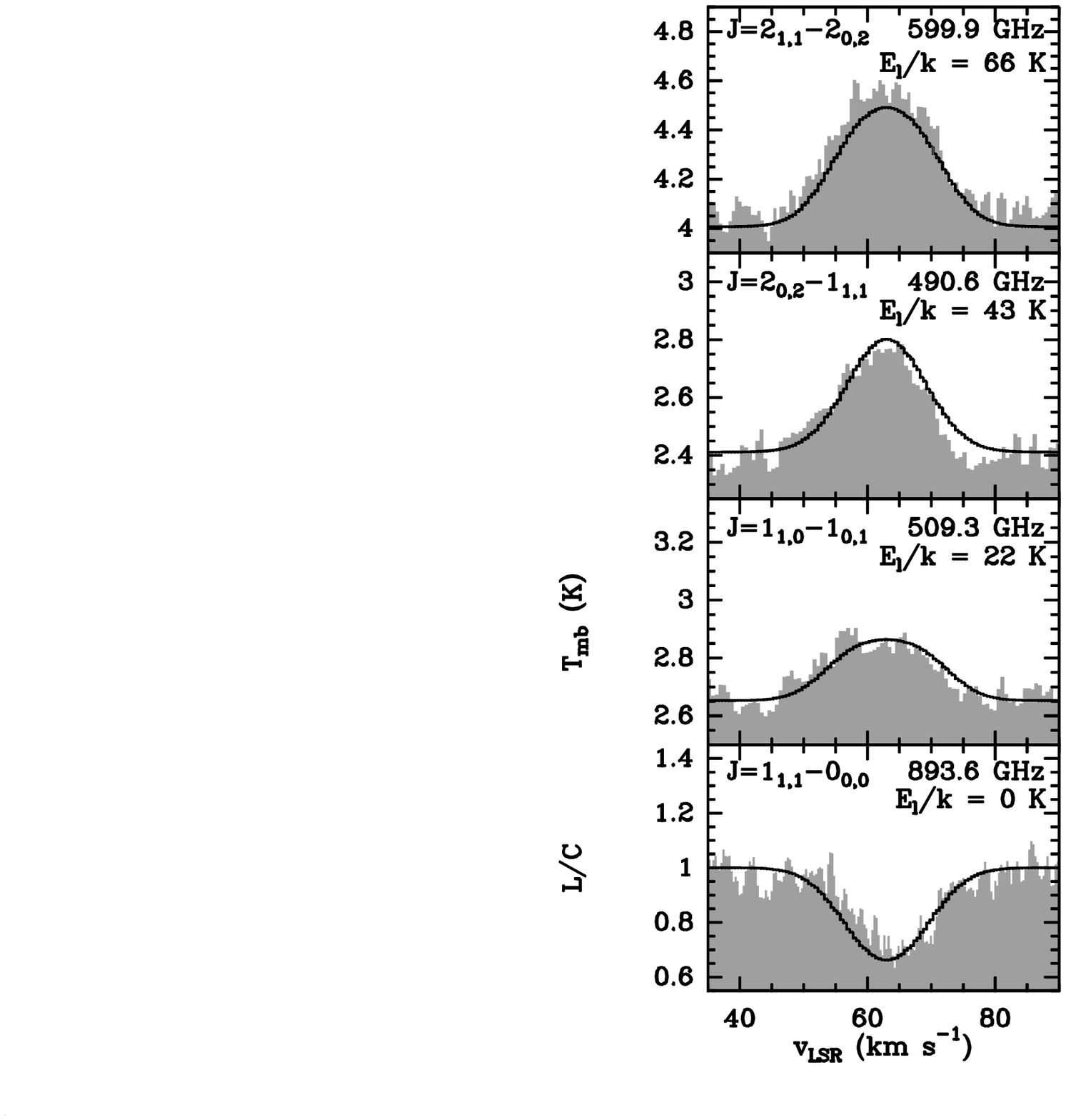}
  \caption{Grey: the top three panels show the HDO emission features detected
with HIFI, the bottom panel shows the ground-state absorption feature detected with the CSO. Our best-fit model, described in \S~\ref{results}, is overlaid in black. { Note that the absorption line is about 2 km$\cdot$s$^{-1}$ off from the center velocity of the emission lines, which is within the uncertainty of the HIFI calibration in HIPE 2.9.}}\label{fig:models_data}
\end{figure}

Figure~\ref{fig:models_data} shows in grey the three HDO emission features detected
with HIFI, and also the CSO absorption feature at the bottom. Our best-fit model
is overlaid in black. The observed HDO features display a half-power width (HPW) of { 14.5}~km$\cdot$s$^{-1}$, and are centered at v$_{LSR} = 63$~km$\cdot$s$^{-1}$, with the exception of the 894-GHz absorption feature, which is centered at v$_{\rm LSR} = 64.7$~km$\cdot$s$^{-1}$ (see Fig.~\ref{fig:models_data}). { This discrepancy is within the uncertainty on the velocity scale that affects HIPE 2.} In this work we will use the same velocity parameters (v$_{LSR} = 63$~km$\cdot$s$^{-1}$, $\Delta$v$=14.5$~km$\cdot$s$^{-1}$) for all features in the model.

It has already been demonstrated (C03 and references therein)
that a constant HDO abundance across the whole envelope of Sgr~B2(M) cannot
reproduce the observed intensities of features spanning a relatively
wide range of excitation, which means that they arise from different regions within the
cloud. It is reasonable to expect a drop in abundance when the gas and
dust temperature fall below 100~K, causing the water content
of the gas to freeze out onto the dust grains. 

\begin{table}[htdp]
\caption{{ Observed and modeled intensities for all the features in our dataset (see Table~\ref{tab:lines}).}}
\begin{center}
\begin{tabular}{rcccc}
\hline
\hline
   & J$_K$  &  $\nu$ &   T$_{\rm mb, obs}$   &   T$_{\rm mb, mod}$ \\
   &                                                    & (GHz)  &             (K)                     &     (K)                            \\
\hline   
1 & $3_{1,2} - 3_{1,3}$                &  481.8  &  $<0.2^{\star}$              & 1$\times 10^{-2}$      \\
2 & $2_{0,2}  - 1_{1,1}$               &  490.6  & 0.4                                &  0.4                               \\
3 & $1_{1,0}  - 1_{0,1}$               &   509.2  &  0.2                              &  0.2                               \\
4 & $2_{2,0} - 3_{0,3}$                &    537.8 &  $<0.1^{\star}$            &   2$\times 10^{-4}$     \\
5 & $2_{1,1} - 2_{0,2}$                &    599.9 &   0.5                            &  0.5                                 \\
\hline
6 &  $1_{1,0} - 1_{1,1}$               &  80.6      &  0.1                             & 4$\times 10^{-2}$       \\
7 & $3_{1,2}  - 2_{2,1}$               &   225.9   &   0.4                            & 0.4                                 \\
8 & $2_{1,1}  - 2_{1,2}$               &   241.6   &   5.5$^{\star\star}$    & 1.7$^{\star\star}$        \\
9 & $1_{0,1}  - 0_ {0,0}$              &   464.9   &    $<0.7^{\star}$          & 0.6                                  \\
10 & $1_{1,1}  - 0_{0,0}$             &   893.6   &    0.7$^{\star\star\star}$   & 0.7              \\
\hline
\end{tabular}
\end{center}
$^{\star}$ = upper limit ($3\times \sigma_{\rm rms}$)\\
$^{\star\star}$ = integrated area (K$\cdot$km$\cdot$s$^{-1}$)\\
$^{\star\star\star}$ = line-to-continuum ratio
\label{tab:model_results}
\end{table}%

With the current model,
a simultaneous fit of all the { HIFI and CSO transitions (see Fig.~\ref{fig:HDO_levels})} is obtained assuming
that [HDO] = $1.5\times10^{-9}$ when T$_{gas} > 100$~K, and [HDO] =
$1.3\times10^{-11}$ when T$_{gas} < 100$~K.  However,  this two-step abundance model underestimates the intensity of the { mm-wavelength features}.

\begin{figure}
   \centering
   \includegraphics[angle=-90,width=9cm]{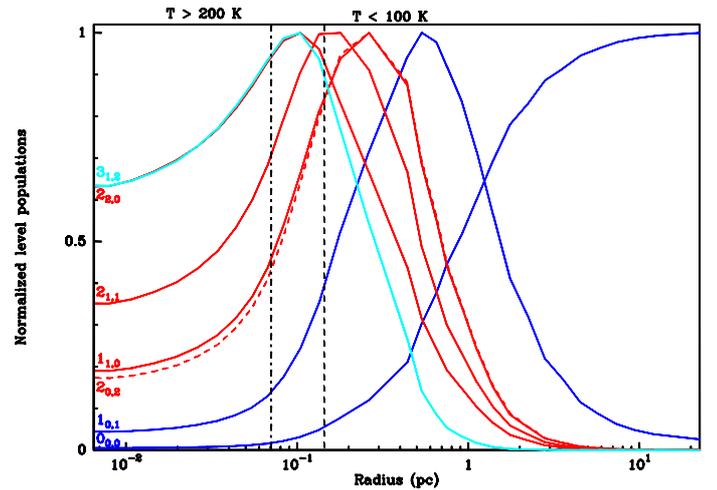}
   \caption{{ Normalized fractional level population computed by RATRAN, as a function of the distance from the center of the cloud, for the HDO transitions shown in red, dark blue and light blue in Fig.~\ref{fig:HDO_levels} (the population of level $2_{0,2}$ is indicated by a dashed red curve, to avoid confusion with level $1_{1,0}$). The levels are \emph{upper-energy}, except for the ground-state 893-GHz absorption line, for which the relevant level population is that of the  \emph{lower-energy} level.}}\label{fig:frac_pop}
\end{figure}

Figure~\ref{fig:frac_pop} shows the normalized populations of the HDO levels of the relevant features in our dataset as calculated by RATRAN. The bulk of the emission for the J=$3_{1,2}-2_{2,1}$ 226-GHz transition is produced in the inner hot-core region, where T$> 200$~K. The other features though arise predominantly in the inner-envelope/outer-core region, with temperatures close to or lower than 100~K.  In order to reproduce the intensities of { the $3_{1,2}-2_{2,1}$ transition as well, we need to introduce a rise in HDO abundance to about $3.5\times10^{-9}$, where  T$_{gas} > 200$~K. 

Our model very precisely reproduces 8 out of the 10 transitions in our sample (see Tab.~\ref{tab:model_results}). On the other hand, the $1_{1,0} - 1_{1,1}$  and the $2_{1,1}  - 2_{1,2}$ transitions, at 80.6 and 241.6~GHz respectively, are underestimated by about a factor 3. Although we cannot completely rule out  contamination by other species, especially at 3-mm wavelength, where the contribution of unidentified features to the spectrum is about $50\%$ \citep{Belloche2008}, we do not think it realistic that these two features are blended with other, weaker ones to the point of appearing three times stronger than they are.  The model's failure to reproduce these lines may be due to the assumption of spherical symmetry, which does not capture all the details of the actual source structure.}

\section{Discussion}

With the \textit{Herschel} lines discussed in this paper, we now possess information about the HDO abundance throughout the envelope (Fig.~\ref{fig:frac_pop}).  The physical structure model we employed is based on the structure used by \citet{Comito2003}, refined by the study of \citet{Rolffs2010}. We note that we can fit 8 out of 10 lines using this model and a very simple abundance distribution: a low abundance ($1.3\times 10^{-11})$ below 100~K, a medium abundance ($1.5\times 10^{-9}$) at temperatures between 100 and 200~K, and a high abundance ($3.5\times 10^{-9}$) at temperatures above 200~K. This is in contrast to the simpler two-step abundance structure in C03, where no constraints on the HDO abundance from the inner envelope/outer core were available. The exact boundaries of these abundance changes are not well determined, and while evaporation of ice produces a real discontinuity, other processes do not.  As mentioned in Sect.~\ref{intro}, no process but ice evaporation can produce HDO in noticeable amounts.  In the hot gas phase, this molecule will be formed relative to water with the same abundance of deuterium relative to hydrogen, $1.7\times 10^{-6}$ in the Galactic center \citep{Lubowich2000} which, even if all the oxygen were tied up in water, would correspond to an abundance of only a few times $10^{-10}$. This is almost two orders of magnitude lower than what is observed in the hot gas.  A possible explanation for the abundance gradient in gas at T~$>100$~K could be that the ice evaporation is not complete at 100~K, and that the gas-phase HDO abundance gradually increases until a temperature of about 200~K is reached; or that in the inner, denser regions, simply more water (and therefore HDO) has formed on the ice mantles.

To make use of the full information contained in this study, as described in Sect.~\ref{intro}, we need to complement it with the study of H$_2$O and its isotopologues, and then determine the [HDO]/[H$_2$O] ratio throughout the envelope.  The water modeling will not be as straightforward, because the higher opacities will make the H$_2$O transitions much more sensitive to deviations from the spherical symmetry. Moreover, water will be produced in outflows (within the core) and shocks (in the outer envelope). However, the comparison will provide necessary clues to the study of the development of the Sgr B2(M) envelope.

\begin{acknowledgements}
This paper is dedicated to the memory of Oliver Siebertz (1967-2010).

HIFI has been designed and built by a consortium of institutes and university departments from across 
Europe, Canada and the United States under the leadership of SRON Netherlands Institute for Space
Research, Groningen, The Netherlands and with major contributions from Germany, France and the US. 
Consortium members are: Canada: CSA, U.Waterloo; France: CESR, LAB, LERMA,  IRAM; Germany: 
KOSMA, MPIfR, MPS; Ireland, NUI Maynooth; Italy: ASI, IFSI-INAF, Osservatorio Astrofisico di Arcetri- 
INAF; Netherlands: SRON, TUD; Poland: CAMK, CBK; Spain: Observatorio Astron�mico Nacional (IGN), 
Centro de Astrobiolog�a (CSIC-INTA). Sweden:  Chalmers University of Technology - MC2, RSS \& GARD; 
Onsala Space Observatory; Swedish National Space Board, Stockholm University - Stockholm Observatory; 
Switzerland: ETH Zurich, FHNW; USA: Caltech, JPL, NHSC.
Support for this work was provided by NASA through an award issued by JPL/Caltech.
CSO is supported by the NSF, award AST-0540882.
\end{acknowledgements}


\end{document}